\documentstyle[aas2pp4,amssym,epsfig]{article}

\righthead{Host Galaxy Disks}
\lefthead{Hunt, L.K. et al.}

\begin{document}

\title{The Disks of Galaxies with Seyfert and Starburst Nuclei:
II. Near-Infrared Structural Properties}

\author{ L. K. Hunt}
\affil{ C. A. I. S. M. I. - C. N. R. \\
Largo E. Fermi 5, I-50125 Firenze, Italy\\
Electronic mail: hunt@arcetri.astro.it}

\author { M. A. Malkan}
\affil{ University of California \\
Department of Astronomy, 405 Hilgard Ave., \\ 
Los Angeles, CA, U.S.A. \\
Electronic mail: malkan@bonnie.astro.ucla.edu}

\author{G. Moriondo}
\affil{ Universit\`a di Firenze \\
Istituto di Astronomia, Largo E. Fermi 5, I-50125 Firenze, Italy \\
Electronic mail:  gmorio@arcetri.astro.it} 

\and

\author { M. Salvati}
\affil{ Osservatorio Astrofisico di Arcetri \\
Largo E. Fermi 5, I-50125 Firenze, Italy \\
Electronic mail:  salvati@arcetri.astro.it} 


\begin{abstract}
We have derived the near-infrared structural components of a sample of Seyfert and
starburst (SBN) host galaxies by fitting the images of Hunt et al.
(\cite{hunt}) with a new two-dimensional decomposition algorithm.
An analysis of the fitted parameters shows that
Seyfert 1 and SBN bulges resemble normal early-type bulges in structure and color, 
with $(J-K)^c_b$ about 0.1~mag redder than disk $(J-K)^c_d$.
Seyfert 2 bulges, instead, are bluer than normal with $(J-K)^c_b\ \sim\ (J-K)^c_d$.
Seyfert disks (especially Type 1), but not those of SBNs, 
are abnormally bright (in surface brightness), 
significantly more so than even the brightest normal disks.
Seyfert disks are also compact, but
similar to those in normal early-type spirals.
For a given mass,
Seyferts and especially SBNs are abnormally rich in neutral hydrogen, 
and there is strong, albeit indirect, evidence for lower mass-to-light ($M/L$) ratios in
Seyfert and SBN {\it disks}, but for normal $M/L$s in their bulges.
In Seyferts and SBNs, HI mass fractions and $M/L$ ratios are anticorrelated,
and we attribute the high gas mass fractions and
low $M/L$s in SBNs and several Seyferts to ongoing star formation.
Such abundant gas in Seyferts would be expected to inhibit bar formation,
which may explain why active galaxies are not always barred.
\end{abstract}


\bigskip\bigskip
\noindent
{\large\it Accepted for publication in ApJ.}

\section{Introduction}

While for many years, work on Seyfert activity was focussed
on the properties of the active nucleus (AGN),
more and more studies of Seyfert activity
are now aimed at characterizing the host galaxy.
Such a trend is mainly motivated by the notion that Seyfert activity,
like violent star formation, needs to be maintained by a
reliable supply of fuel, the source of which is thought to be in the 
galactic disk.
Recent theoretical work, for example,
has shown that non-axisymmetric potentials, such as bars,  
are effective movers of gas into the central potential well, thus
providing a mechanism for feeding the Seyfert nucleus
(Norman \cite{norman}; Shlosman, Begelman, \& Frank \cite{shlosman-frank};
Barnes \& Hernquist \cite{barnes}).
Such non-axisymmetric potentials could be induced by galactic
encounters (e.g., Hernquist \cite{hernquist}), or by cool disks, 
unstable to perturbations (Heller \& Shlosman \cite{heller-shlosman}).

The putative overabundance of bars in Seyferts is, however, 
still a subject of controversy.
Some statistical work suggests that Seyfert nuclei are more prevalent in 
barred galaxies (Arsenault \cite{arsenault}), but
other authors conclude that the frequency
of bars in Seyfert hosts is the same as that in normal spirals
(Moles, Marquez, \& Perez \cite{moles}).
Observational studies of Seyferts have found
no excess of bars relative to normal galaxies
(McLeod \& Rieke \cite{mcleod}; Ho, Filippenko, \& Sargent \cite{ho:1997};
Mulchaey \& Regan \cite{mulchaey}).
Moreover,
when the centers of Seyfert galaxies are examined at high resolution,
there is still no evidence that either inner bars or
double/multiple nuclei are unusually common (Malkan, Gorjian, \& Tam \cite{malkan}).

Interactions in Seyfert galaxies are subject to a similar controversy. 
While many studies find a connection between interactions
and Seyfert activity
(Dahari \cite{dahari}; Keel et al. \cite{keel-kennicutt};
Fuentes-Williams \& Stocke \cite{fuentes}; Rafanelli et al. \cite{rafanelli}),
AGNs are also found in isolated galaxies
(Moles et al. \cite{moles}; Keel \cite{keel:1996}).
Moreover,
it appears that Seyfert activity is virtually absent in very disrupted systems 
(Keel et al. \cite{keel-kennicutt}; Bushouse \cite{bushouse}).

Recent work on the connection between environmental factors and 
nuclear activity has also concentrated on the properties of the bulge
(e.g., Nelson \& Whittle \cite{nelson-whittle1}).
Seyfert nuclei tend to reside preferentially in early-type spirals
(e.g., Moles et al. \cite{moles}), and intuitively the bulge
would be expected to be connected with the properties 
of the central gravitational potential well. 
If massive compact objects (black holes) lie at the heart of 
Seyfert activity (c.f., Terlevich et al. \cite{terlevich}), 
the importance of the bulge can be appreciated by considering the
correlation of putative black hole mass with the mass of the bulge 
(Kormendy \& Richstone \cite{kormendy-richstone}).
The relative dominance of the bulge in a galaxy may also be
connected with the size and strength of a bar, and with the locations of
resonances where gas piles up and star formation ensues 
(Elmegreen \& Elmegreen \cite{elmegreen}; Combes \& Elmegreen \cite{combes}).

In this paper, we study both structural components of Seyfert host
galaxies by decomposing the two-dimensional images into a bulge and a disk. 
In an earlier paper (Hunt et al. \cite{hunt}, hereafter Paper I), near-infrared
(NIR) images of galaxies with Seyfert and starburst nuclei (SBN) were
used to quantify the host galaxies in terms of stellar populations. 
We found that both Seyfert and SBN host galaxies have 
NIR colors that reflect, in the mean, normal late-type stellar populations.
We also found non-axisymmetric red colors primarily in SBNs and blue
colors in Seyfert 2s, which could be interpreted as a (possibly evolutionary?)
sequence from SBNs, to Type 2, to Type 1 Seyferts. 
Here, we have decomposed the NIR images in Paper I into bulge and disk 
components, and have compared their structural parameters and component colors
with those from normal spirals.
In $\S$2, we briefly describe the Seyfert and SBN samples introduced in
Paper I, together with 
the control samples of normal spirals to which we applied the same
decomposition algorithm.
The image decomposition is described in $\S$3, and 
the structural parameters of Seyfert and SBN bulges and disks are given
in $\S$4, and compared with those of normal galaxies in $\S$5.
Implications of our results are discussed in $\S$6.
We defer analysis of the non-axisymmetric 
features of Seyfert galaxies to a much larger sample 
(Hunt et al., in preparation), and discussion of the nuclear amplitudes
and colors to a future paper. 
As in Paper I, we use a Hubble constant $H_0\,=\,$75~km\,s$^{-1}$Mpc$^{-1}$.

\section{The Samples}\label{samples}\protect

The Seyferts studied here and in Paper I are part of the CfA sample
(Huchra \& Burg \cite{huchra:burg}), and the SBNs were selected from
the Markarian lists as identified and compiled by Balzano (\cite{balzano})
and Mazzarella \& Balzano (\cite{mazzarella:balzano}).
Both samples are magnitude-limited, and we imposed a redshift constraint
of $z \ge$~0.015 to ensure compatibility with detector field-of-view.
The selection criteria and observations are described in detail in Paper I
where $J$-, $H$-, and $K$-band images of 26 Seyfert 1s, Seyfert 2s, and
SBNs are presented.
One of the SBNs, Mrk~732, is, in reality, a  Seyfert 1
(Osterbrock \& Pogge \cite{osterbrock:pogge}), and has therefore been
included as a Seyfert in the statistical analysis.

Besides these ``active'' objects,
two additional normal spiral samples for which we have NIR images have also been
considered in this paper.
The first is selected from the normal spiral galaxies described in 
de Jong \& van der Kruit (\cite{dejong:vdkruit}) and in de Jong 
(\cite{dejong:thesis}), and is a minimum-diameter-limited sample.
This data set was also used as a control sample in Paper I, and
as mentioned there, we used only those objects with well-defined spiral 
types between Sa and Sc, and the same selection criteria as for the 
program samples were applied, that is to say $z \ge$~0.015.
The subset of the de Jong galaxies (hereafter referred to as ``Sc's'')
considered here and in Paper I is dominated by later spiral types; 
only two galaxies have a morphological type earlier than Sb, and 
most are clustered around Sc.
One of the ``Sc's'' is Mrk~545 (UGC 89, NGC 23), a SBN,  
and was eliminated from consideration in the statistical comparison.

The second normal galaxy sample comprises the early-type spirals studied in Moriondo, 
Giovanardi, \& Hunt (\cite{moriondo:1}, \cite{moriondo:2}; hereafter MGH1, MGH2).
These galaxies are taken mainly from the work of  Rubin and 
collaborators (Rubin et al. \cite{rubin}) who selected Sa's from
Sandage \& Tammann  (\cite{sandage}).
They are relatively luminous 
systems with dominant bulges; the observing sample reported in MGH1 contains
only those galaxies with apparent diameter $\lesssim$ 4~arcmin.
The morphological type spread in this sample is tight with only
one galaxy (NGC 5879) as late as Sbc, and the remaining objects are Sab or 
earlier. 
The Moriondo (hereafter referred to as ``Sa's'')
and de Jong samples together span the entire range of spiral types,
with early-types well represented for comparison with the Seyferts.

We anticipate here the sample medians for $K$-band galaxy total luminosity: 
the medians are, in descending order,
$-24.5$ (Sy 1's), $-24.3$ (Sc's), $-24.2$ (Sa's), $-23.9$ (Sy 2's), and $-23.8$ (SBNs).
The samples are similar in luminosity,
and, moreover, the spread for each sample is also similar
and relatively small, about 3~mag. 
Therefore, we do not expect to find luminosity-dependent biases 
in the derived structural parameters.

\section{Galaxy Decomposition \label{decomp}\protect}

Previous studies of Seyfert galaxies that examined the host galaxy
bulge and disk relied on the decomposition of
one-dimensional (1D) surface brightness (SB) profiles
(Yee \cite{yee}; MacKenty \cite{mackenty}; Kotilainen et al. \cite{kotilainen};
Danese et al. \cite{danese})\footnote{Yee and MacKenty
applied a single-component disk model.}. 
Such techniques generally fail to provide reliable galaxy structural parameters 
(Byun \& Freeman \cite{byun-freeman}).
This can be understood by considering the general case of a
system inclined relative to the plane of the sky.
When a galaxy is inclined,
projected bulge and disk components have quite different shapes, and
their sum is not well-represented by elliptical isophotes.
Hence, even if the galaxy would be well-described by the 
simple two-component bulge$+$disk model, 
it is difficult if not impossible for 1D profile-fitting to 
estimate correctly the true parameters (Burstein \cite{burstein}).

We have therefore adopted a technique which fits two-dimensional (2D) 
surface brightness distributions as
the sum of a bulge, a disk, and an unresolved point source.
The bulge, assumed to be an oblate rotational ellipsoid
coaxial with the disk, is modelled with a generalized exponential
(S\` ersic \cite{sersic}, Sparks \cite{sparks}):
\begin{equation}
\label{bulge_b}\protect
I_b\,(x,y)\ =\ I_e\exp
\left\{-\alpha_n \left[\left(\frac{1}{r_e}\sqrt{x^2+\frac{y^2}
{(1-\epsilon_b)^2}}\ \right)^{1/n}-1\right]\right\}
\end{equation}
Bulge index $n\,=\,$4 corresponds to the ``standard'' $R^{1/4}$ law.
$I_e$ ($\mu_e$ in magnitude units)
and $r_e$ are effective (half-light) surface brightness and radius,
$\epsilon_b$ is the apparent bulge ellipticity,
and $\alpha_n$ is a constant relating
the effective brightness and radius to the exponential values 
(see MGH1).
The bulge apparent eccentricity $e_b$ is related to the
intrinsic eccentricity $e_b'$ by:
\begin{equation}
\label{ecce}\protect
e_b = e_b'\sin\,i \;\; .
\end{equation}
The disk, assumed to be thin, is modelled with a simple exponential:
\begin{equation}
\label{disk_b}\protect
I_d\,(x,y)\ =\ I_d(0)\exp\left[-\frac{1}{r_d}\sqrt{x^2+
\frac{y^2}{\cos^2 \,i}}\ \right] \; .
\end{equation}
$I_d(0)$ ($\mu_d$ in magnitude units)
and $r_d$ are the central surface brightness
and exponential folding length, respectively, and $i$ is the
system inclination.
The unresolved nuclear source is assumed to be a delta function 
with amplitude $I_n(0)$ ($m_n$ in magnitude units).

Structural parameters were determined by fitting the model,
convolved with a circular Gaussian PSF, 
to the data in flux units using a $\chi^2$ minimization. 
Weights were assigned to each pixel according to the 
photon noise expected in background-limited 
performance, behavior that was verified observationally.
For each galaxy, the bulge was fit with four values of the 
exponent $n$: 1, 2, 3, and 4.
Because the field-of-view of IRCAM1 is relatively small,
there are rarely stars in the same image as the galaxy. 
Hence, the seeing width $\sigma$ was determined by fitting a circular
Gaussian to standard stars taken close in time to the program
objects.
The galaxy center was determined by applying Gaussian fits to the images,
and the position angle of the galaxy on the sky was 
determined when possible from 
NED\footnote{The NASA/IPAC Extragalactic Database (NED) 
is operated by the Jet Propulsion
Laboratory, California Institute of Technology, under contract with the U.S.
National Aeronautics and Space Administration.},
or otherwise from the images themselves.

The algorithm fits three parameters for the bulge
($I_e$, $r_e$, and $\epsilon_b$), three for the disk
($I_d(0)$, $r_d$, $i$), and one for the nucleus ($I_n(0)$).
Because we saw no evidence for changes in scale length with
wavelength, and because simulations demonstrated that parameters
are generally more reliable when more than one wavelength is
fit simultaneously, 
each galaxy was fit using all of the available
images in different filters.
The bulge and disk scale lengths,
$r_e$ and $r_d$, are therefore assumed to be the same for all wavebands, 
as are bulge ellipticity $e_b$ and system inclination $i$.
For three bands, then, there are 13 free parameters in the fit.
Bulge index $n$ is not free to vary, as mentioned above,
so that the value of $n$ that gave the best overall fit, $n_{\it best}$,
was chosen a posteriori. 
The fitting routine is described in detail in MGH1.

One feature of the decomposition used here bears special mention.
Before fitting, the galaxy image is folded about its major
and minor axes to generate only one quadrant. 
Such a procedure is possible because of the axial symmetry of the model.
The uncertainty for each point is calculated as the maximum of 
the photon noise, as mentioned above, and the deviations from symmetry
measured in the folding process.
When bars, oval distortions, or other asymmetric structures are not coincident
with the line of nodes, such deviations are much larger than the photon
statistical uncertainties, and those regions are therefore fit with lower weight than
more symmetric ones.
Hence, sensitivity to non-axisymmetric structures is much reduced,
relative to a more conventional fitting technique, making the algorithm
less subject to confusion between bars or lenses and bulges or disks.

\subsection{Numerical Simulations \label{simul}\protect}

The problem of separating the nucleus in active galaxies from the bulge
does not have a simple solution.
To assess how well our algorithm succeeds in this task,
in addition to the simulations reported in MGH1,
we fit a series of synthetic galaxies with an active nucleus, bulge, and 
disk components. 
To better approximate the NIR observing conditions, 
we added background and photon noise typical of those in our images.
$J$, $H$, and $K$ images were synthesized for each simulation, 
convolved with an appropriate PSF, then fit 
using the same procedure as for the program galaxies.

Our main concern was the accuracy of bulge parameters when $r_e$
is comparable with the seeing $\sigma$.
Hence, for all synthetic galaxies,
we fixed the disk, the nucleus, and the seeing width, 
and varied bulge $n$, $e_b$, $r_e$, and $\mu_e$.
The nucleus and the disk were chosen to have
the same properties as the sample medians 
(disk with $r_d$ = 6~arcsec, $\mu_d(K)$ = 17.0, $J-K$ = 0.91, $H-K$ = 0.20,
$i$ = 45$^\circ$; nucleus with $m_n(K)$ = 13.0, $(J-K)_n$ = 1.5,
$(H-K)_n$ = 0.5).
The seeing FWHM was fixed to 1.6, 1.4, and 1.3~arcsec,
in $J$, $H$, and $K$, respectively.

It turns out that the most difficult parameter to determine
correctly is bulge $n_{\it best}$.
This is the parameter that, in effect, describes the {\it shape} of the
bulge; fixing $n$ is the same as assuming homologous shapes for all bulges. 
If this quantity can be unambiguously determined,
then the fitting algorithm returns the correct 
parameters of all components within the uncertainties.
In general, large-$n$ bulges are equally well fit by either $n_{\it best}$~=~3,~4,
while their small-$n$ (1, 2) counterparts give correct (unambiguous) $n_{\it best}$. 
Spherical bulges with $r_e\,\sim\,\sigma_{seeing}$ are equally well fit by any $n$.

Perhaps the most important result of our simulations is
that in all cases the disk parameters are well determined, 
as are the nuclear amplitudes and colors.
For all simulations, independently of the correct $n_{\it best}$,
we found that the fitted disk $r_d$ and the nuclear $m_n$ (and colors)
reproduce those of the synthetic galaxies to within a few percent.
Disk $\mu_d$ (and colors) are also fit to within a few percent, except
for large-$r_e$, large-$n$ bulges where 
the fitted disk surface brightness can be wrong by as 
much as 0.3~mag\,arcsec$^{-2}$, although colors remain accurate.

\section{Results }

The major-axis cut extracted from the image together with the
best-fit decomposition is shown for each program galaxy
in Fig. \ref{fig:profiles}.
The results of our 2D decompositions are reported in
Table \ref{tbl:parameters}. 
If the best-fitting bulge exponent $n_{\it best}$ can be well-determined,
the bulge parameters can be considered reliable.
``Well-determined'' means that there is a unique lowest value of
$\chi^2$ for a specific value of bulge exponent $n$.
For one-third of each of the samples analyzed (Sy~1, Sy~2, SBN, Sa, Sc),
this is not true and bulge $n_{\it best}$ is ``ill-determined'';
thus for these objects
the bulge parameters are less reliable than they would be otherwise
(and are marked with a colon in Table \ref{tbl:parameters}).

\begin{table}
\dummytable\label{tbl:parameters}
\end{table}

We have checked that components are not mis-identified by the
decomposition algorithm. 
For example, bars may be mistaken for either highly-flattened bulges 
or disks. 
Since roughly half of each of the ``active'' samples are barred, 
at least as noted in de Vaucouleurs et al. (\cite{rc3}) -- RC3 (see Paper I), 
this presents a potential problem.
Bars are generally characterized by a large aspect ratio, and 
in particular, if bars are interpreted as disks by the fitting algorithm,
the fitted inclinations should be larger than the ``nominal'' ones
(those given by RC3 or measured from the outer $J$-band isophotes). 
The mean ratio of nominal to fitted inclinations for
the active samples as a whole is 0.95 with a scatter of 0.5;
for the Seyfert 1 sample alone the ratio is 0.89.
It appears that the fitted inclinations do not differ
systematically from the ``nominal'' ones.
We conclude, albeit tentatively, that the ``disks'' given by the 
decomposition really are the photometric structures normally called disks.
Mis-identification of bars for bulges is more difficult to check,
since we have no way of determining, a priori, the
intrinsic ellipticity of a given bulge.
Nevertheless, only Mrk~545 (SBN) has a fitted bulge with $e_b$ (0.45) which
exceeds expected ranges of oblate ellipsoids (Mihalas \& Binney \cite{mihalas}).

\begin{figure}
\centerline{\epsfig{figure=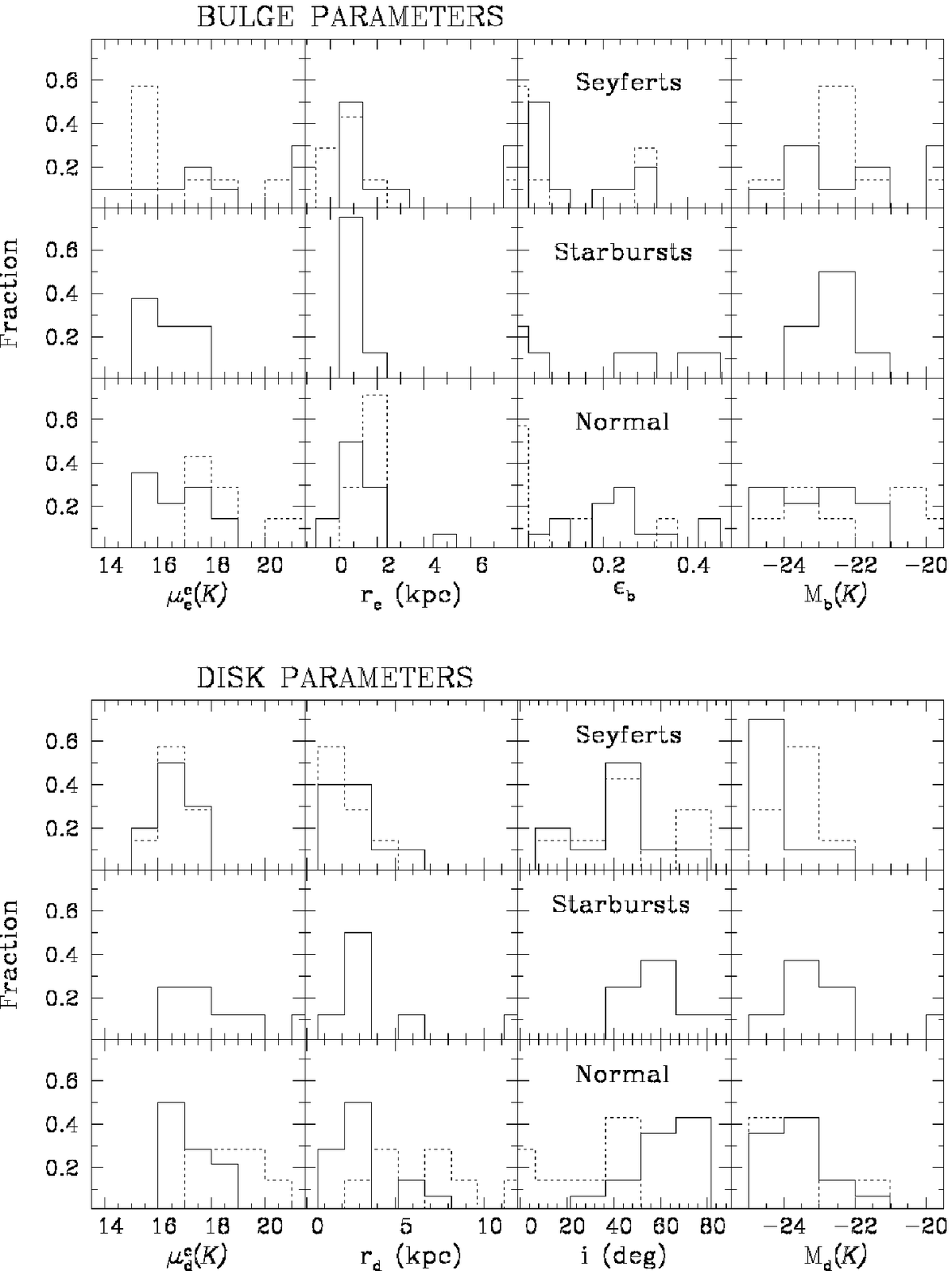,width=12cm}}
\figcaption[fig.his.ps]{Observed distributions of fitted
parameters for each component as a function of galaxy activity class.
Illustrated for bulge (top panel) and disk (bottom panel) are the
$K$-band surface brightnesses and absolute luminosities, scale lengths,
and ellipticities/inclinations.
The top panel for each component shows Type 1 Seyferts as solid lines, and Type 2
as short-dashed lines;
the middle panel shows SBNs; 
the bottom panel shows Sa's as solid lines, and Sc's as short-dashed
lines.
\label{fig:histograms}
}
\end{figure}

The fitted parameters for each of the samples are shown as histograms
in Fig. \ref{fig:histograms}.
Table \ref{tbl:median} gives the individual sample medians for all 
of the fitted parameters.

\begin{deluxetable}{cclllll}

\tablecolumns{7}
\tableheadfrac{0.1}
\small
\tablewidth{0pt}
\tablenum{2}
\tablecaption{Sample Medians of Structural Parameters \label{tbl:median}}
\tablehead{
& & \multicolumn{3}{c}{Active} & \multicolumn{2}{c}{Normal Spirals} \\
\multicolumn{1}{c}{Component} & 
\multicolumn{1}{c}{Parameter} &
\multicolumn{1}{c}{Seyfert 1} &
\multicolumn{1}{c}{Seyfert 2} &
\multicolumn{1}{c}{SBN} &
\multicolumn{1}{c}{Sa's} &
\multicolumn{1}{c}{Sc's} \\
\colhead{(1)} & \colhead{(2)} & \colhead{(3)} & \colhead{(4)} &
\colhead{(5)} & \colhead{(6)} & \colhead{(7)} }
\startdata
       & $n_{best}$   & \phs\phn 2 & \phs\phn 3 & \phs\phn 1 & \phs\phn 3 & \phs\phn 2 \nl
       & $\epsilon_b$ & \phs\phn 0.089 & \phs\phn 0.018 & \phs\phn 0.231 & \phs\phn 0.238 & \phs\phn 0.015  \nl
       & $r_e$~(kpc) & \phs\phn 0.98 & \phs\phn 0.59 & \phs\phn 0.75 & \phs\phn 1.12 & \phs\phn 1.88 \nl
Bulge  & $\mu_e^c(K)$ & \phs 16.2 & \phs 15.7 & \phs 16.3 & \phs 16.7 & \phs 18.1 \nl
       & $(J-K)^c_b$ & \phs\phn 1.04 & \phs\phn 0.87 & \phs\phn 0.99 & \phs\phn 1.07 & \multicolumn{1}{c}{\nodata} \nl
       & $(H-K)^c_b$ & \phs\phn 0.17 & \phs\phn 0.18 & \phs\phn 0.21 & \multicolumn{1}{c}{\nodata} & \phs\phn 0.25 \nl
       & $M_b(K)$ & $-$23.5 & $-$22.6 & $-$22.7 & $-$23.0 & $-$22.7 \nl
\tablevspace{5pt}
\hline
\tablevspace{5pt}
       & $r_d$~(kpc) & \phs\phn 2.5 & \phs\phn 2.1 & \phs\phn 2.8 & \phs\phn 3.2 & \phs\phn 7.7 \nl
       & $\mu_d^c(K)$ & \phs 16.6 &  \phs 16.7 &  \phs 17.3 &  \phs 17.2 &  \phs 18.3 \nl
Disk   & $(J-K)^c_d$ & \phs\phn 0.94 & \phs\phn 0.93 & \phs\phn 0.87 & \phs\phn 0.90 & \multicolumn{1}{c}{\nodata} \nl
       & $(H-K)^c_d$ & \phs\phn 0.21 & \phs\phn 0.20 & \phs\phn 0.25 & \multicolumn{1}{c}{\nodata} & \phs\phn 0.18\nl
       & $M_d(K)$ & $-$24.2 & $-$23.6 & $-$23.3 & $-$23.5 & $-$23.8 \nl
\enddata

\end{deluxetable}

Typical Seyfert and SBN bulges are similar,
being relatively spherical, compact, and bright,
with $r_e\,\sim\,$1~kpc and $\mu_e^c(K)\,\sim\,$16~mag\,arcsec$^{-2}$.
They tend to differ only in 
the bulge exponent $n_{\it best}$, or ``shape parameter'', with Seyferts
having $n_{\it best}\,\sim\,$2--3, and SBNs having $n_{\it best}\,\sim\,$1.
Seyfert 1 bulges appear to be the most luminous, with $M_b(K)\,\sim\,-23.5$.

The typical Seyfert disk is also compact and bright, 
with $r_d\,\sim\,$2~kpc, and $\mu_d^c(K)$ of around 16.7~mag\,arcsec$^{-2}$.
SBN disks are slightly more extended and faint, and appear similar
to early-type disks, with
$r_d\,\sim\,$2.8~kpc, and $\mu_d^c(K)\,\sim\,17.3$~mag\,arcsec$^{-2}$.
Seyfert disks are also luminous, especially those of Seyfert 1s,
with a median $M_d(K)\,\sim\,-24.2$.

\subsection{Bulge and Disk Colors}

As foreseen in Paper I, disk colors $(J-K)^c_d$ and $(H-K)^c_d$ of the active samples
are similar to those in normal spirals. 
Furthermore, as in most spirals (MGH1),
bulge $(J-K)^c_b$ of all the galaxies {\it except the Type 2 Seyferts}
are $\sim\, 0.1$~mag redder than disk $(J-K)^c_d$.
In particular, 
Sa bulges are significantly (97\% one-tailed) redder than their disks. 
On the other hand, Seyfert 2 bulges tend to be blue with $(J-K)^c_b\,\sim\,0.9$, 
comparable to the disk color.
This bulge color in Seyfert 2s is
significantly (95\% one-tailed) bluer than $(J-K)^c_b$ in normal bulges. 

\begin{figure}
\centerline{\epsfig{figure=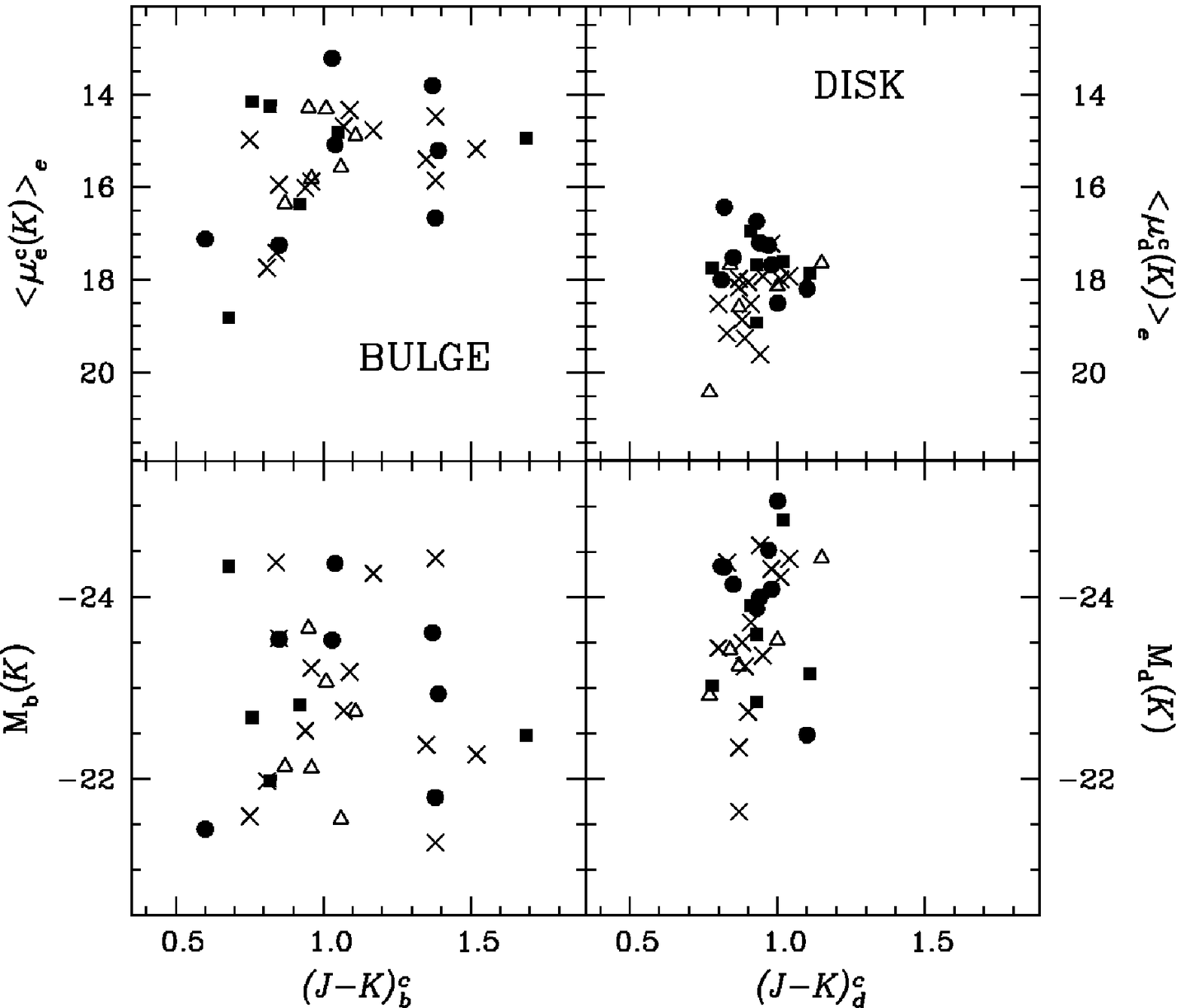,width=12cm}}
\figcaption[jk.ps]{Plots of mean surface brightness $<\mu^c(K)>_e$ and
luminosity $M(K)$ versus $(J-K)^c$.
Bulge parameters are shown in the left panels, and disks on the right.
Seyfert 1s are denoted by filled circles, Seyfert 2s by filled squares,
SBNs by open triangles, Sa's by $\times$, and Sc's by $+$.
\label{fig:jk}
}
\end{figure}

Bulge and disk $J-K$ colors plotted versus $\langle\mu^c(K)\rangle_e$ and $M(K)$ are shown
in Fig. \ref{fig:jk}.
There may be a slight trend for bright bulges to have redder colors, but the
anomalous bulge colors [$(J-K)^c_b \gtrsim$~1.2] do not conform to the trend.
The tightness of the disk colors is evident in the right panels of Fig. \ref{fig:jk}.
Neither bulge nor disk colors depend on the component absolute luminosity.

We have compared the fitted bulge and disk colors to the inner and
outer disk colors derived in Paper I.
The inner disk colors in Paper I were defined from averaging the
elliptical profiles from 4~arcsec to 3~kpc, and the outer from
3~kpc to the noise limit of the profile.
While the 1D outer colors correlate well with the 2D-fitted
disk colors, there is very little connection between the 1D inner colors
and those of the 2D-fitted bulge.
Given the small bulge $r_e$ in our samples, together with the
brightness of the disk, the disk contribution becomes comparable
to that of the bulge well within the 3-kpc limit, 
and thus ``contaminates'' the inner light.
It appears that to properly measure colors of the bulge, especially
the relatively compact ones observed in Seyferts and SBNs (and Sa's),
the effects of the disk and nucleus need to be removed, a task that is 
best accomplished with (2D) decomposition. 

\section{Statistical Comparisons \label{stats:comp}\protect}

The Kolmogorov-Smirnov (K-S) test 
evaluates the probability that two observed cumulative distributions
have been drawn from the same parent population.
We have applied the K-S test to each pair of samples for each parameter,
on the basis of the distributions shown in Fig. \ref{fig:histograms}.
The results are illustrated schematically in Fig. \ref{salvati.diagram}.
As mentioned in the caption,
we distinguish between low significance level (between 90\% and 95\% shown 
in lower case), moderate (between 95 and 99\% in upper case), and 
high ($>$\,99\% in bold face).
Although only those differences at 95\% or greater should be considered
significant, certain trends emerge when the 
lower-probability differences are taken into account.

\subsection{Comparative Summary}

Bulges of Seyfert, SBN, and normal spirals are similar to one another in most 
photometric properties.
The most salient difference
is between the bulges of normal early- and late-type spirals, as
Sa bulges on average are $>\,$1\,$K$-mag\,arcsec$^{-2}$
brighter than those in Sc's.
This result confirms the work of de Jong
(\cite{dejong:iii}), who found that early-type bulges are characterized
by brighter $\mu_e$, independently of the type of bulge parameterization
($n_{\it best}$).

Disk properties differ substantially among the different spiral and active types. 
Fig. \ref{salvati.diagram} shows that the Sc disks differ
significantly from those of early-type and active spirals:
they are measurably more tenuous and extended than their early-type counterparts.
Indeed, there appears to be a progression from large, tenuous late-type spiral disks
($r_d$\,$>$\,5\,kpc and $\mu_d^c(K)$\,$\sim$\,18),
to brighter and more compact early-types 
($r_d$\,$\sim$\,3\,kpc and $\mu_d^c(K)$\,$\sim$\,17).
While part of this behavior may result from selection effects\footnote{The Sc's
are taken from a minimum-diameter sample which is expected to be less biased
against low-SB objects than are magnitude-limited samples
(McGaugh et al. \cite{mcgaugh-bothun}).}, 
MGH1 also found that early-type spiral disks are more than 1\,mag\,arcsec$^{-2}$
brighter than late-type disks (Giovanardi \& Hunt \cite{giova:1988}), similar
to the trend of early- and late-type bulges.

The K-S tests also show that
Seyfert disks are {\it significantly brighter} even than Sa disks.
Seyfert disks are also compact, but not significantly more so than 
the Sa's, and like early-type disks, they are significantly more compact than 
those in late-type spirals.
Although the Seyferts, SBNs, and Sa's are selected from
magnitude-limited samples which tend to be biased against low surface-brightness
objects (McGaugh et al. \cite{mcgaugh-bothun}),
they are also ``maximum-diameter'' selected
(the Sa's explicitly so, and the Seyferts and SBNs indirectly
so because of the redshift constraint).
Furthermore,
as mentioned in $\S$\,\ref{samples}, the median luminosities and spreads of the
different samples are comparable.
The similarity of the selection criteria and luminosities among the Seyferts,
SBNs, and Sa's
would therefore suggest that any bias is not acting ``differentially'' to 
affect our results. 

\subsection{Correlations of Surface Brightness and Scale Length}

We investigate here whether bulge and disk parameters for Seyferts and SBN's obey relations
similar to those in normal spirals.
Correlations between surface brightness and scale length in elliptical
galaxies and of bulges and disks in spiral galaxies have been known for
some time 
(Kormendy \cite{kormendy}; Hoessel \& Schneider \cite{hoessel};
Djorgovski \& Davis \cite{djor:davis};
Kent \cite{kent:1985}; Kodaira, Watanabe, \& Okamura \cite{kodaira}).
These two parameters, in fact, constitute an almost face-on view 
of the ``fundamental plane'' (FP) for bulges and ellipticals 
(e.g., Kormendy \& Djorgovski \cite{kormendy:djor}, and references
therein).

Figure \ref{fig:fpk} shows plots of bulge and disk SB against
respective effective scale lengths\footnote{The disk exponential
folding length $r_d$ has been converted to the half-light radius,
so that the bulge and disk plots have the same units.}.
The Seyfert and SBN parameters are shown in the upper panels, and the
normal galaxies in the lower panels.
Also shown in the lower left panel are bulges taken from
Bender, Burstein, \& Faber (\cite{b2f}) and
Andredakis, Peletier, \& Balcells (\cite{apb}) converted to the $K$ band according
to the prescription given by Andredakis et al.
The dotted lines in the left panels show 
the best-fit line to our normal Sa and Sc bulges\footnote{The slope is
calculated as the ordinary least-squares (OLS) bisector
given by Isobe et al. (\cite{isobe:1}), and 
the intercept is defined by forcing the regression to pass through
the barycenter of the data points.}.
The best-fit slope of 2.90 corresponds to $r_e\,\propto\,I_e^{-0.86}$,
very similar to what is found by Bender et al. (\cite{b2f}) and
earlier work (e.g., Kormendy \& Djorgovski \cite{kormendy:djor}).
Moreover, the intercept relative to the $B$-band conversion gives
a mean $B-K$ color of 3.8, which is a rather normal color for these systems. 
The upper left panel in Fig. \ref{fig:fpk} shows that the Seyfert and SBN
bulges follow very closely the trend defined by normal spiral bulges
and ellipticals.

The situation changes for the disks.
We have fit our normal Sa and Sc disks in the same way as for the 
bulges\footnote{The low-luminosity systems in the Sc sample, UGC~628
and UGC~12845, have been omitted from the fit. Given that in dwarf
galaxies the trends of scale length $r$ and $\mu$ are
contrary to those in luminous systems where $r$ increases with fading $\mu$
(Binggeli \& Cameron \cite{binggeli}), this should be legitimate.},
and find a best-fit slope of 2.94, similar to the bulge value, and
corresponding to $r_e\,\propto\,I_e^{-0.85}$, identical to that found
for ellipticals.
This regression is shown, with the normalization appropriate for
normal disks, as the lower dotted
line in the right panels of Fig. \ref{fig:fpk};
the upper dotted line shows the analogous trend for the bulges, 
repeated from the left panels. 
The figure shows that, for a given disk $r_e$, Seyfert disks
have a mean SB 0.9\,$K$\,mag\,arcsec$^{-2}$ {\it brighter}
than those in even early-type spirals, comparable to the typical 
surface brightnesses of the bulge. 
This appears to be another confirmation of the abnormally bright
surface brightness of Seyfert disks.

\subsection{Mass-to-Light Ratios }

To complete our discussion of the properties of Seyfert host
galaxy components, we have collected kinematic data from the literature in the form
of central velocity dispersions (Nelson \& Whittle \cite{nelson-whittle1};  
Prugniel \& Simien \cite{prugniel:simien}) and neutral hydrogen velocity widths (RC3). 
These data have been used to derive global mass-to-light ($M/L$) ratios for 
the galaxies, and relative $M/L$ ratios for their bulges.
Global galaxy $M/L$s rely on an effective radius $r_{\it eff}$ for the
galaxy as a whole, obtained by analytically integrating the bulge$+$disk model.
Hydrogen velocity widths have been corrected for inclination (using
fitted values) and turbulence.
As in Burstein et al. (\cite{burstein:1997}),
we have used the expression for $\kappa_3$ to calculate relative bulge $M/L$s 
($\kappa_3 \equiv (\log \,\sigma_c^2 - \log \,I_e - \log \,r_e)/\sqrt{3}$), and
$r_{\it eff}$ to calculate global masses.

It turns out that effective $K$-band $M/L$ ratios of Seyfert galaxies tend to be 
slightly smaller than those of normal early-type spirals 
(median Seyfert $M/L \,=\,0.7$, median Sa $\,=\,0.9$ in solar units). 
If the actively star-forming Sa's are eliminated from the 
median\footnote{Hereafter, this subset of Sa's will be denoted as ``true Sa's''.}
(see MGH1), then
the Seyfert $M/L$ ratios result almost a factor of two smaller 
(median true Sa $\,=\,1.2$);
the difference in the sample means (0.8 for Sy's vs. 1.5 for true Sa's) is 
significant at the 96\% (one-tailed) level.
These $M/L$ ratios are plotted in the upper panels of Fig. \ref{fig:ml2}.

We would argue that the lower $M/L$ ratio of Seyferts is due to their {\it disks}.
While Nelson \& Whittle (\cite{nelson-whittle2}) attribute an offset in
Seyfert bulge luminosity versus velocity dispersion 
regressions (e.g., Faber-Jackson) to reduced {\it bulge} $M/L$, 
a Faber-Jackson plot for the bulges in our
samples reveals no significant offset between Seyferts and Sa's.
A possible explanation of the discrepancy lies in their determination of bulge
luminosity, which relies on mean bulge-to-disk ratios as a function of morphological
type (e.g., Simien \& de Vaucouleurs \cite{simien}).
For the Seyferts (especially Type 1) in our sample such corrections, 
on average, {\it overestimate} the bulge by 0.6~mag; 
instead, the normal early-type bulges are well determined with the standard
correction, having a mean error of 0.1~mag. 
For a given morphological type, therefore, 
B/D ratios for Seyferts appear to be somewhat smaller than those of normal spirals.
Indeed, the Sy 1's, similar to the Sa's in almost every way, have significantly more
luminous disks (see Fig. \ref{salvati.diagram}), 
an effect which probably results from their high surface
brightnesses as shown in the upper right panels of Figs. \ref{fig:jk} and \ref{fig:fpk}.
The mean difference of 0.6~mag that we find between Seyferts and normal 
early-type spirals is perhaps fortuitously similar to Nelson \& Whittle's 
mean offset of 0.7~mag, but may be an explanation for the disagreement.

We have also placed Seyferts, SBNs, and Sa's in the fundamental plane
according to the formalism of Burstein et al. (\cite{burstein:1997}, see also
Bender et al. \cite{b2f}). 
Figure \ref{fig:kappa} shows various projections of the FP together with
the $B$-band relations defining the FP in the $\kappa_1/\kappa_3$
projection (shown as a solid line), and the ``zone of exclusion'' in $\kappa_1/\kappa_2$.
As before, our data have been converted from $K$ to $B$ according to
Andredakis et al. (\cite{apb}).
In these projections of the FP at a given mass, the $M/L$s of Seyfert {\it bulges} 
(upper left panel) are very similar to those in normal spiral bulges. 
In contrast, our $K$-band data for global $M/L$s (upper right panel) 
do not strictly conform to the FP as defined in $B$ by Burstein et al. 
(\cite{burstein:1997}). 
However, because of our small sample sizes (further reduced by the available
kinematic data), we have not determined a slope, but rather
fixed it to the canonical one, and for each sample adjusted the intercept 
(``offset'') in a least-squares sense. 
The offsets for Sy's (shown as dotted line) and true Sa's (dashed line)
differ significantly (99.9\% --one-tailed-- level),
again with Sy's having global $M/L$s almost a factor of two lower.
This result, though, is not independent of 
the smaller Seyfert $M/L$ ratio previously discussed, since it depends on 
the same input quantities.
What we have shown here is that while true Sa's conform to the FP,
Seyferts do not; they tend to have, in the mean, lower global $M/L$s.

We therefore find no evidence for a systematically lower $M/L$ ratio in Seyfert bulges,
relative to normal early-type spirals, but rather for smaller {\it global} $M/L$ ratios 
in Seyfert galaxies. 
In the absence of spatially-resolved kinematic data, it is not straightforward
to determine absolute $M/L$ ratios of the bulge and disk, 
but indirect arguments seem to suggest that the Seyfert disk, not the bulge, 
is the galaxy component with an anomalously low $M/L$ ratio.
This point will be discussed further in the next section.

\subsection{Neutral Hydrogen Content \label{hi}\protect}

We have calculated two quantities, both evaluated with $r_{\it eff}$,
to measure the neutral hydrogen content in
our samples: the mean HI surface density $\sigma_{HI}$, and the HI mass fraction
$M_{HI}/M_{\it eff}$. 
Seyfert, SBN, and Sc median $\sigma_{HI}$
are typical of spiral types Sb or later
(Roberts \& Haynes \cite{roberts:haynes}), while  
Sa's show median $\sigma_{HI}$ three or four times lower, 
consistent with normal early-type spirals (Eder et al. \cite{eder}).
In terms of neutral hydrogen mass fraction,
SBN's have the highest median 
$M_{HI}/M_{\it eff}$ of all the samples ($\sim$\ 40-50\%), with
Seyferts following at 18\% (14\% for Type 1's, 20\% Type 2's).
Sa's at 4\% are consistent with normal early-type spirals
(Roberts \& Haynes \cite{roberts:haynes}; Broeils \& Rhee \cite{broeils-rhee}).
On the basis of both criteria, therefore, we conclude that Seyferts,
although structurally very similar to Sa's, are much richer in neutral gas.
Indeed,
even normal very late-type spirals are observed to have 
$M_{HI}/M_{\it eff}$ $\lesssim$\,10\,--12\,\% (Roberts \& Haynes \cite{roberts:haynes}), 
so that the Seyferts, and
especially SBNs, appear to have abnormally high neutral gas fractions,
independently of their morphological type.
If the molecular gas were taken into account, the work of Maiolino
et al. (\cite{maiolino}) shows that we could even be underestimating the
total gas content in these systems by a factor of two or more.

We also find the HI mass fraction $M_{HI}/M_{\it eff}$ in Seyferts and SBNs
to be anticorrelated with effective $M/L$ ratio.
This anticorrelation is shown in the upper middle panel of Fig. \ref{fig:ml2}
where lower $M/L$ is seen together with higher $M_{HI}/M_{\it eff}$, but
the trend is amplified by the use of $M_{\it eff}$ in both plotted variables. 
To assess the impact of this on the correlation,
we have performed Monte Carlo experiments that reproduce the statistical
characteristics of the combined Sy+SBN sample (in terms of mean and spread 
in $L_K$, $M_{\it eff}$, $M_{HI}$, and the correlations between them). 
Results show (via the Fisher $z$ test, see Bulmer \cite{bulmer}) that
the anticorrelation between $M_{HI}/M_{\it eff}$ and $M/L$ is significant at the 
96\% (two-tailed) level (2$\sigma$),
independently of underlying correlations 
and those induced by correlated measurement uncertainties.
The behavior of the Sa's, on the other hand, is well-reproduced by the
Monte Carlo experiments, appearing to depend solely upon the underlying biases.

The dependence of gas content on $M/L$ is also shown in the upper
right panel of Fig. \ref{fig:ml2}, where $M/L$ is plotted against $\sigma_{HI}$. 
The HI surface brightness is independent of distance, and the correlation
should be more free of systematic effects than $M_{HI}/M_{\it eff}$ vs. $M/L$;
$\sigma_{HI}$ and $M/L$ are again significantly correlated in Seyferts and 
SBNs at the 2$\sigma$ level.
Nevertheless, true HI surface brightness is only very
crudely estimated by the values reported here, and
we would argue that the more physically significant variable is 
$M_{HI}/M_{\it eff}$.
Even for anomalously high gas mass fractions,
spatially resolved HI maps are needed to determine if Seyferts and SBNs
really do have, for a given morphology, higher-than-average HI surface
density.

We also find in Seyferts and SBNs a correlation 
($\gtrsim$\ 95\% two-tailed) between $\mu_d(K)$ and HI mass fraction, 
in the sense that fainter $\mu_d(K)$ implies higher $M_{HI}/M_{\it eff}$.
This trend, shown in the lower left panel of Fig. \ref{fig:ml2}, 
is displaced from normal spirals, and the clear segregation  
between active and normal spirals is indicated by the diagonal dotted line.
Since HI tends to be associated with the stellar disk, these are yet
more indications that the low $M/L$ ratios we and others find in
Seyfert galaxies are a property of the disk, and not of the bulge.

\section{Summary, Discussion, and Speculation \protect\label{discussion}}

We summarize here what has emerged from our investigation, and
then speculate about what the results might imply.

\begin{figure}
\centerline{\epsfig{figure=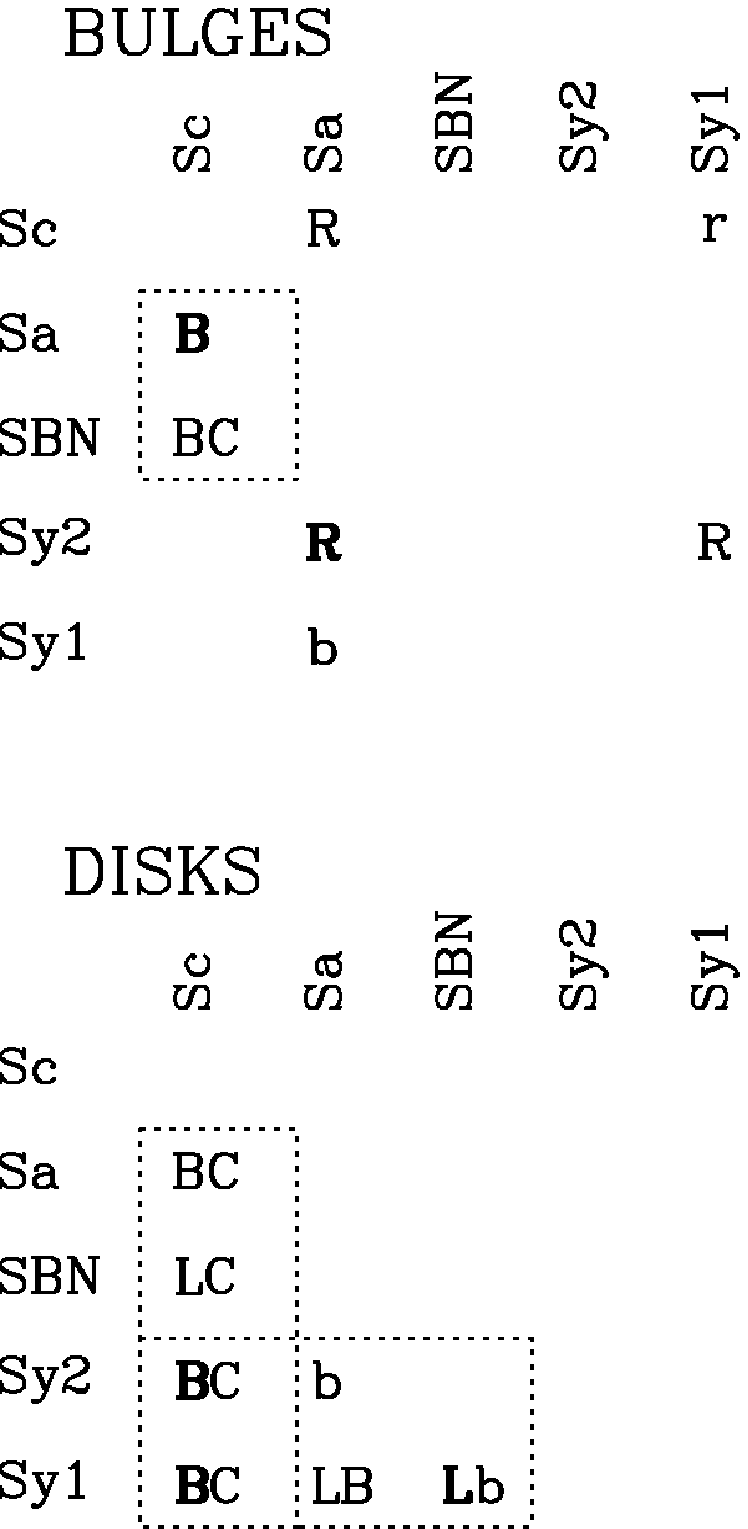,width=7cm}}
\figcaption[salvati.diagram.ps]{A graphic summary of the 
statistical analysis described in $\S$~\ref{stats:comp}.
The upper panel refers to bulge properties, the lower one to disks. The
letters indicate the quantity for which a significant difference was
found between the two given subsamples: B stands for surface brightness,
R for roundness of bulges (i.e. rounder bulges have smaller $\epsilon_b$),
C for compactness (i.e. more compact components have smaller physical
scale length), and L for luminosity.
The sign of the 
difference (i.e., which sample has the largest value of a given parameter)
is from left to above, in the sense that for a given comparison,
the sample towards the bottom and left has a smaller value of the parameter
than the sample to the top and right.
The significance of the difference
is given by the size of the letter: lower case means a significance level
between 90 and 95\%, upper case between 95 and 99\%, and bold face greater
than 99\%. For instance, considering only Sc's and Sa's in the upper panel,
we see that Sc bulges are rounder than Sa's at between 90 and 95\% confidence
level, whereas Sa bulges have higher surface brightnesses than Sc's at
more than 99\%.
Dotted lines are given to better delineate the
significant differences between samples.
\label{salvati.diagram}
}
\end{figure}

\begin{figure}
\centerline{\epsfig{figure=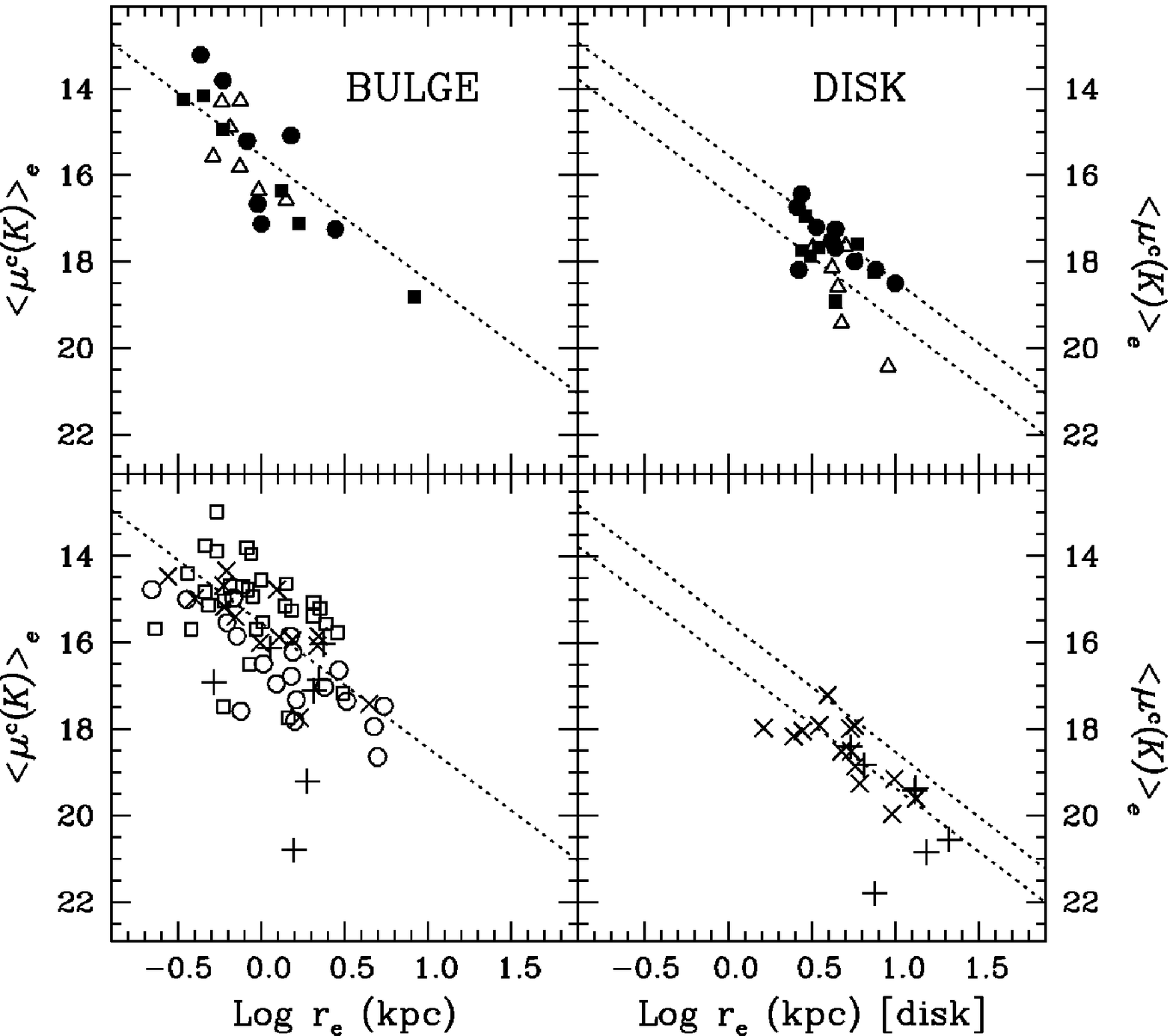,width=12cm}}
\figcaption[fpk.ps]{Plots of mean surface brightness $<\mu^c(K)>_e$
versus log\,$r_e$.
Bulge parameters are shown in the left panels, and disks on the right;
active spirals are shown in the upper panels, and normal spirals in the lower ones.
Disk exponential folding length $r_d$ has been converted to the half-light
effective radius $r_e$, so that bulge and disk plots have the same units.
As in Fig. \ref{fig:jk},
Seyfert 1s are shown as filled circles, Seyfert 2s as filled squares,
SBNs as open triangles, Sa's as $\times$, and Sc's as $+$.
In the lower left panel,
bulges from Bender et al. (\cite{b2f}) and from Andredakis et al. (\cite{apb})
are shown as open circles and open squares, respectively, converted to $K$ as
described in the text.
The regressions shown as dotted lines in the left panels are the best
fits to our normal bulges as defined in the text.
The right panels show the bulge regression as the upper line, and the
best fit to the normal disks as the lower line;
they have similar slope, but
are offset vertically by roughly 0.9\,mag\,arcsec$^{-2}$.
\label{fig:fpk}
}
\end{figure}

\begin{figure}
\centerline{\epsfig{figure=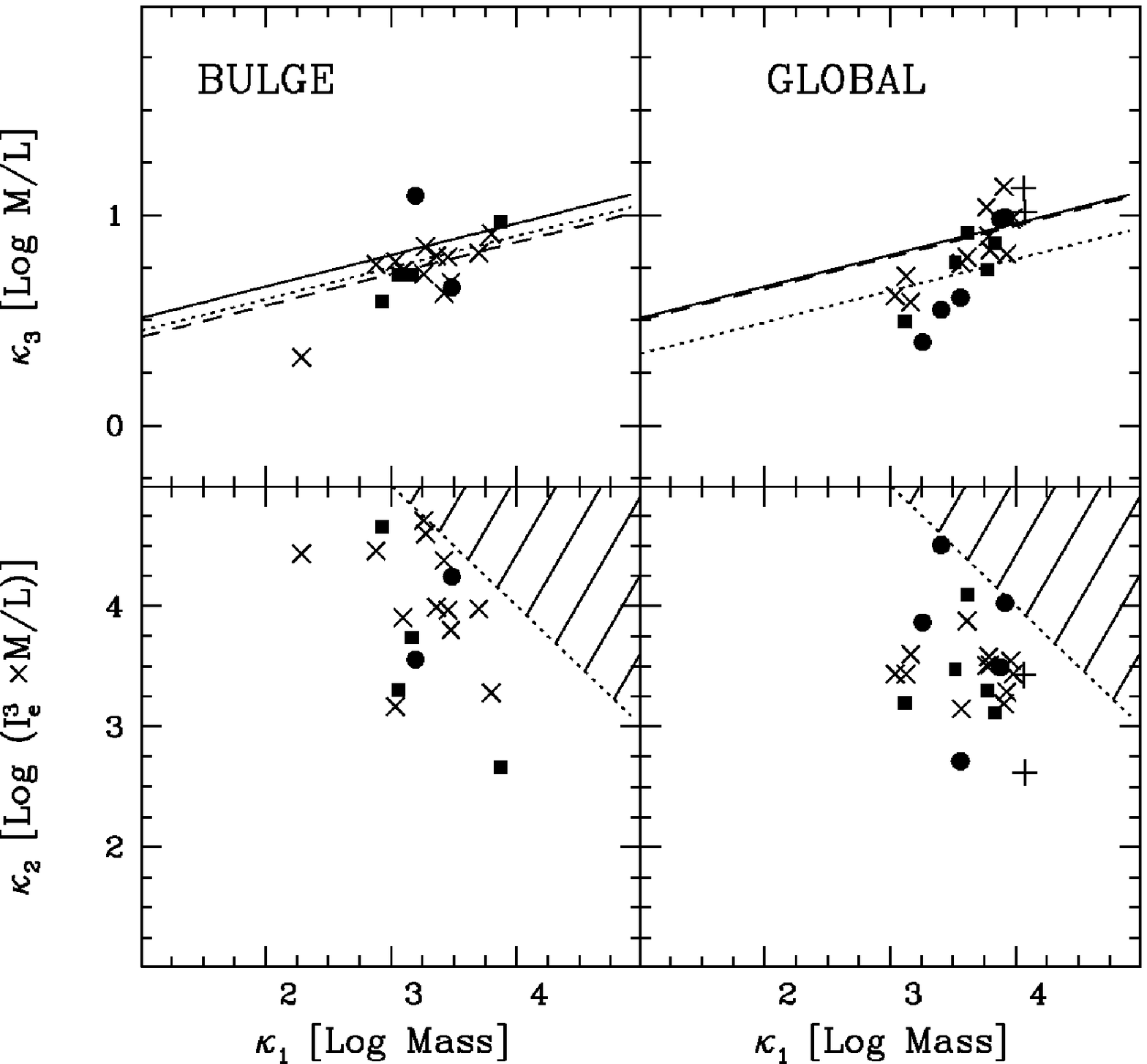,height=12cm}}
\figcaption[kappa.ps]{$\kappa_1/\kappa_2$ and $\kappa_1/\kappa_3$ projections
of the FP as defined in Bender et al. (\cite{b2f}).
$\kappa_1 \equiv (\log \,\sigma_c^2 + \log \,r_e)/\sqrt{2}$;
$\kappa_2 \equiv (\log \,\sigma_c^2 + 2\,\log \,I_e - \log \,r_e)/\sqrt{6}$;
$\kappa_3 \equiv (\log \,\sigma_c^2 - \log \,I_e - \log \,r_e)/\sqrt{3}$.
Bulge parameters are shown in the left panels, 
with $I_e$ and $r_e$ defined in the $K$ band from our decompositions,
and $\sigma_c$ taken from Nelson \& Whittle (\cite{nelson-whittle1}).
Global effective parameters are shown in the right panels, with $K$-band $I_{\it eff}$ 
and $r_{\it eff}$ derived from the fits as described in the text.
HI line widths are taken from RC3, and corrected as described in Burstein et al.
(\cite{burstein:1997}).
All data have been converted to the $B$ band as described in the text.
Symbols are as in Fig. \ref{fig:jk}.
The upper panels, with $\kappa_1/\kappa_3$ shows the solid line defined
by: $\kappa_3 = 0.15\,\kappa_1 + 0.36$,
and the lower panels with $\kappa_1/\kappa_2$ shows the zone-of-exclusion
defined by: $\kappa_1 + \kappa_2 = 8$.
Offsets as defined in the text
are shown as dashed lines for the Sa's, and as dotted lines for the Sy's.
The lower global (effective) $M/L$ for the Seyferts is evident in the upper right panel. 
\label{fig:kappa}
}
\end{figure}

\begin{figure}
\centerline{\epsfig{figure=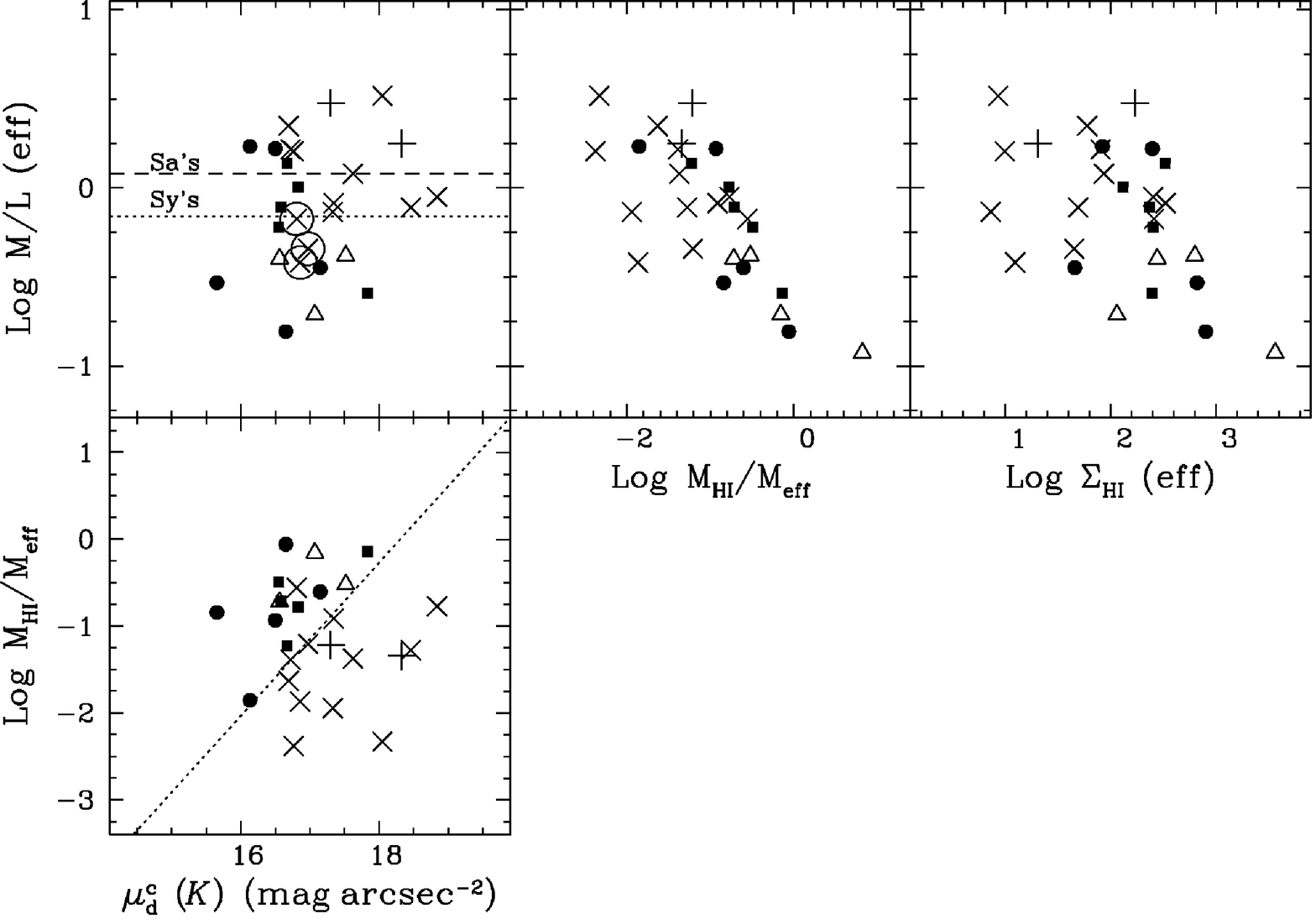,width=14cm,angle=270}}
\figcaption[ml2-4.ps]{Plots of (log) effective $M/L$ versus $\mu^c_d(K)$ (upper left panel);
versus (log) neutral hydrogen mass fraction $M_{HI}/M_{\it eff}$ (upper middle panel);
versus (log) neutral hydrogen surface density $\sigma_{HI}$ (upper right panel);
and $M_{HI}/M_{\it eff}$ versus $\mu^c_d(K)$ (lower panel).
$M/L$ ratios are effective (half-light) calculated as described in the text.
As in Fig. \ref{fig:jk},
Seyfert 1s are shown as filled circles, Seyfert 2s as filled squares,
SBNs as open triangles, Sa's as $\times$, and Sc's as $+$.
The three Sa's in the upper left panel are actively
star-forming Sa's (NGC~3593, 4419, 5879), and are marked with circumscribing circles; 
the median Sa (omitting these) and Seyfert $M/L$ are illustrated with horizontal dotted lines.
The dotted line in the lower left panel is not a fit, but rather indicative
of the segregation of the ``active'' and normal spiral classes.
\label{fig:ml2}
}
\end{figure}

\begin{itemize}
\item
In normal spirals, we confirm the
trends found in earlier studies of bulge and disk parameters with morphological type:
early-type spiral  bulges are brighter ($\mu_e$) than late-types
(de Jong \cite{dejong:iii}),
and Sa disks are more compact ($r_d$) and bright ($\mu_d$) than those in Sc's
(GMH1).
\item
Seyfert and SBN bulges resemble normal bulges in structure and color, with
$(J-K)^c_b$ about 0.1~mag redder than disk $(J-K)^c_d$, except that:
$i)$~Sy 2 bulges tend to be rounder than those in Sy 1's and normal spirals;
$ii)$~Sy 2 bulges have $(J-K)^c_b\ \sim\ (J-K)^c_d$ (see also Paper I).
\item
Seyfert disks (especially Type 1), but not those of SBNs, 
are significantly brighter than early-type disks, 
which are, as mentioned above, brighter than those of Sc's.
Evidence of this result is seen
in the K-S comparision (Fig. \ref{salvati.diagram}) and
in the trends of disk SB and radius (Fig. \ref{fig:fpk}).
Seyfert disks are also compact, but not significantly more so than those in
early-type spirals.  
We also find some suggestion (Fig. \ref{salvati.diagram})
for, at a given morphological type, higher disk {\it luminosity} in Sy 1's 
(hence reduced bulge-to-disk ratios)
which we attribute to their higher disk surface brightness.
\item
Seyferts, and especially SBNs, are very rich in neutral gas:
while neutral gas fractions $M_{HI}/M_{\it eff}$ for normal spirals  
are observed to be $\lesssim$\ 10\%, Seyferts have $M_{HI}/M_{\it eff}$
of around 18\%, and SBNs $\gtrsim$\ 40\%.
This trend is illustrated
in Fig. \ref{fig:ml2}, where the different properties of normal 
and active spirals are evident.
\item
We find strong, albeit indirect, evidence for lower $M/L$ ratios in
Seyfert and SBN {\it disks}, but for normal $M/L$s in their bulges.
Such evidence includes normal (i.e., without offset) Faber-Jackson plots
for Seyfert bulges; an offset in the FP, not for the bulges, but rather
towards lower {\it global} $M/L$ ratios (Fig. \ref{fig:kappa});
and anticorrelations of $M/L$ with what are usually
thought to be a disk quantities, namely HI mass fraction and surface 
density (Fig. \ref{fig:ml2}).
\end{itemize}

First, we find Seyfert disks to be anomalously bright, relatively compact,
and, for their mass, abnormally rich in HI. 
Evolutionary simulations of disk galaxies suggest that such disks might be the 
product of mass transport from the outer regions to the center,
and the ensuing episode(s) of star formation.
Models of galaxies with a stellar bulge, disk, and a gas component show a significant 
central buildup of {\it stars and gas} after 2~Gyr 
(Junqueira \& Combes \cite{junqueira}).
In the Junqueira \& Combes models,
the evolved stellar (and gaseous) distributions show progressively 
brighter central surface brightnesses and smaller effective radii,
similar to those we observe in Seyfert galaxies.
The weak trend (Fig. \ref{fig:ml2}) we find for dimmer $\mu_d(K)$ at higher 
$M_{HI}/M_{\it eff}$ in Seyferts and SBNs 
(and perhaps in normal spirals, see McGaugh \& de Blok \cite{mcgaugh})
may also support this scenario:
the more gas already converted into stars, the lower the gas mass fraction, 
and the brighter the disk. 

Second,
such secular buildup and central concentration of stars
and gas occurs in the simulations {\it even in the absence of a bar}.
The high neutral gas mass fraction in Seyferts (and SBNs) 
may be an explanation for the discordant findings of 
incidence of bars and interactions in Seyfert galaxies
(see Introduction and references therein).
The simulations of Junqueira \& Combes (\cite{junqueira}) correspond
to a gas fraction of 10\%, but, as noted by them,
when gas mass fractions exceed roughly 10\%, the gas can damp
bar instabilities, instead of exciting them
(Shlosman \& Noguchi \cite{shlosman-noguchi}).
The abnormally abundant neutral gas in Seyferts may therefore prevent the
formation of bars in Seyfert disks;
the gas gets funnelled inward anyway because of dynamical friction
(Shlosman \& Noguchi \cite{shlosman-noguchi}) or spiral wave instabilities
(Junqueira \& Combes \cite{junqueira}; Zhang \cite{zhang}).

Third, 
the anticorrelation between $M/L$ and $M_{HI}/M_{\it eff}$ suggests that
SBN and some Seyfert disks have undergone or are undergoing more star 
formation than normal spirals.
For a given age,
$M/L$ ratios measured at wavelengths longward of about 1\,$\mu$m 
decrease with metallicity (Worthey \cite{worthey}), and,
at a given metallicity, increase with age.
According to the Worthey models,
the expected range due to age is 0.5\,dex for ages $\gtrsim$\,2~Gyr, 
and in metallicity less than half that.
Because the $M/L$s observed here anticorrelate with
neutral hydrogen mass fraction, and because their variation 
has an amplitude ($\sim$ 2\,dex) that 
exceeds by far the spread predicted by the Worthey models,
we claim that the $M/L$ ratios indicate a very young age, not extreme metallicity.
If this is true, the low $M/L$ ratios and high
$M_{HI}/M_{\it eff}$ imply that the SBNs and some Seyferts are actively forming stars,
and their stellar populations are relatively young ($<$\,2~Gyr).
Turning again to Fig. \ref{fig:ml2}, we add that among Seyfert and SBNs,
and perhaps separately among normal spirals, the more gas already converted into
stars, the higher the $M/L$, and the older the mean age of the stellar population.

As to differences in star formation histories between Seyfert 1s and 2s,
we are not able, because of our small samples, to make 
conclusive statements.
Although the statistics are very poor, Type 2 Seyferts may have
a higher HI mass fraction than Type 1s, suggesting that star formation 
may have occurred more recently in Seyfert 2s
(Paper I; Oliva et al. \cite{oliva} ). 
The blue disk-like color and roundness of Seyfert 2 bulges are probably relevant here. 
Blue $J-K$ bulge colors should indicate younger age or lower metallicity or both, although
we have very little color leverage to disentangle 
the combined effects of age and metal abundance. 
We can only point out that 
Seyfert 2 bulges are more similar to normal and active spiral {\it disks}, 
than to normal bulges, or to the bulges of Type 1 Seyferts and SBNs.

Finally,
we address the question of an evolutionary link between Seyfert, SBN,
and normal bulges and disks.
Normal early-type spiral bulges and disks are brighter 
and more compact than those of late types.
Since we argued above that bright, compact disks could result
from inward mass transport in Seyferts, we could speculate in a similar vein
that normal early-type disks have evolved, over time, from late-types. 
The central depression in neutral hydrogen commonly observed in early-type spirals
(e.g., van Driel \& van Woerden \cite{vandriel})
would result naturally from such a scenario, since presumably the gas would
be blown out (e.g., Israel \& van Driel \cite{israel}) or
consumed (van Driel \& van Woerden \cite{vandriel})
by violent star formation in the central regions. 
Disk evolution is also compatible with
recent work that suggests that the rate of galaxy evolution depends on disk surface density.
Gas consumption is delayed in low-surface-brightness systems, and brighter
disks are more evolved, having undergone more episodes of star
formation in the past (McGaugh \& de Blok \cite{mcgaugh}).

We speculate further that Seyferts may be extreme cases of
high (initial and present) HI mass fraction coupled with 
high-density, evolved disks at the current epoch.
Because of their high gas content, 
Seyfert disks have continued to undergo episodes of star formation
up to now, consuming the gas, but not yet exhausting it.
Early-type spirals would be those which did not have sufficient gas initally to 
create either a bright Seyfert-like disk or an active nucleus,
and SBNs would not yet have had enough time to build up either one, but may do so 
in the future given the large reservoir of gas available.
The rate of bulge/disk/active nucleus evolution in spirals
would be determined by the gas fraction, the
efficiency and time scale of the inward mass transport, 
and by the rate and efficiency of star formation.

\acknowledgements
We would like to thank Edvige Corbelli, Daniele Galli, and Francesco Palla for
interesting discussions; Carlo Giovanardi for insightful
comments; and an anonymous referee for probing questions that resulted in a
better paper.
This research was partially funded by ASI Grant ARS-96-66.

\figcaption[newfig.ps]{$J$-band major-axis cuts together with
the best-fit decomposition.
The observed and fit $J-K$, $J-H$, and $H-K$ color profiles are
also shown; these are merely the difference of the major-axis cuts
in the various filters.
The major-axis cuts are shown as crosses, and
elliptically-extracted profiles (see Paper I) are shown as open circles.
Error bars indicate the uncertainties used in the 2D decomposition,
and may be dominated by asymmetries as described in the text. 
The top horizontal axis shows galactocentric distance in kpc,
assuming the distances given in Table \ref{tbl:parameters}, and the bottom
galactocentric distance in arcsec.
Bulge $n_{\it best}$ is shown in the upper right corner below the object name.
The three components are shown in the upper panel as long dashed lines,
and their sum in all panels as short dashed lines.
The fits take into account the
different seeing widths in the different bands, but the observed profiles
have not been rebinned.
The profile of Mrk~993, the most highly inclined galaxy in our samples,
shows a good
example of the difference between elliptically-averaged profiles
and the major-axis cut.
Not only do the surface brightnesses differ substantially in the
central region, but also the colors ($J-K$, $J-H$).
\label{fig:profiles}
} 

\onecolumn

\begin{figure}
\centerline{\epsfig{figure=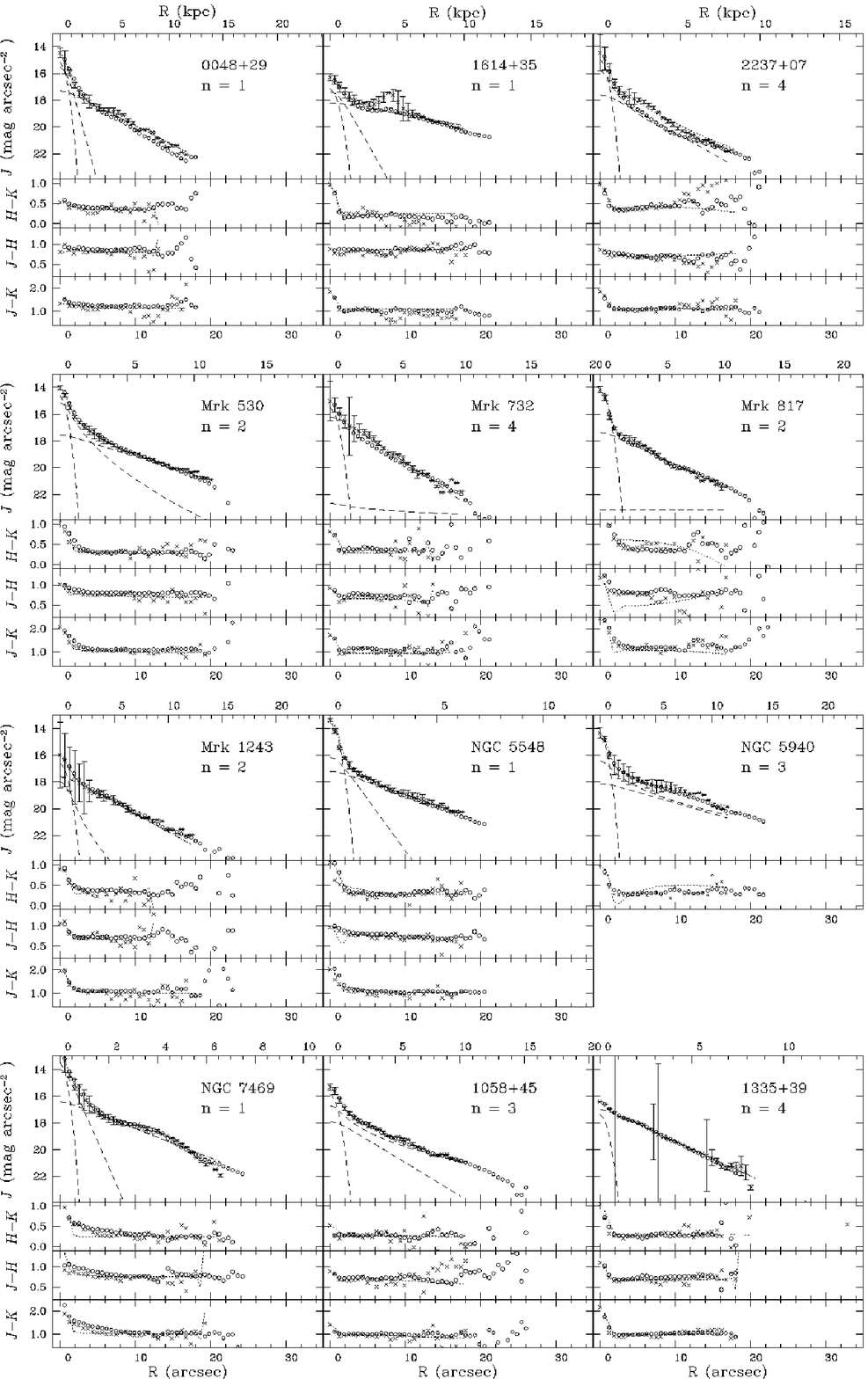,width=13cm}}
\medskip
\centerline{\bf Fig. 1a}
\end{figure}

\begin{figure*}
\centerline{\epsfig{figure=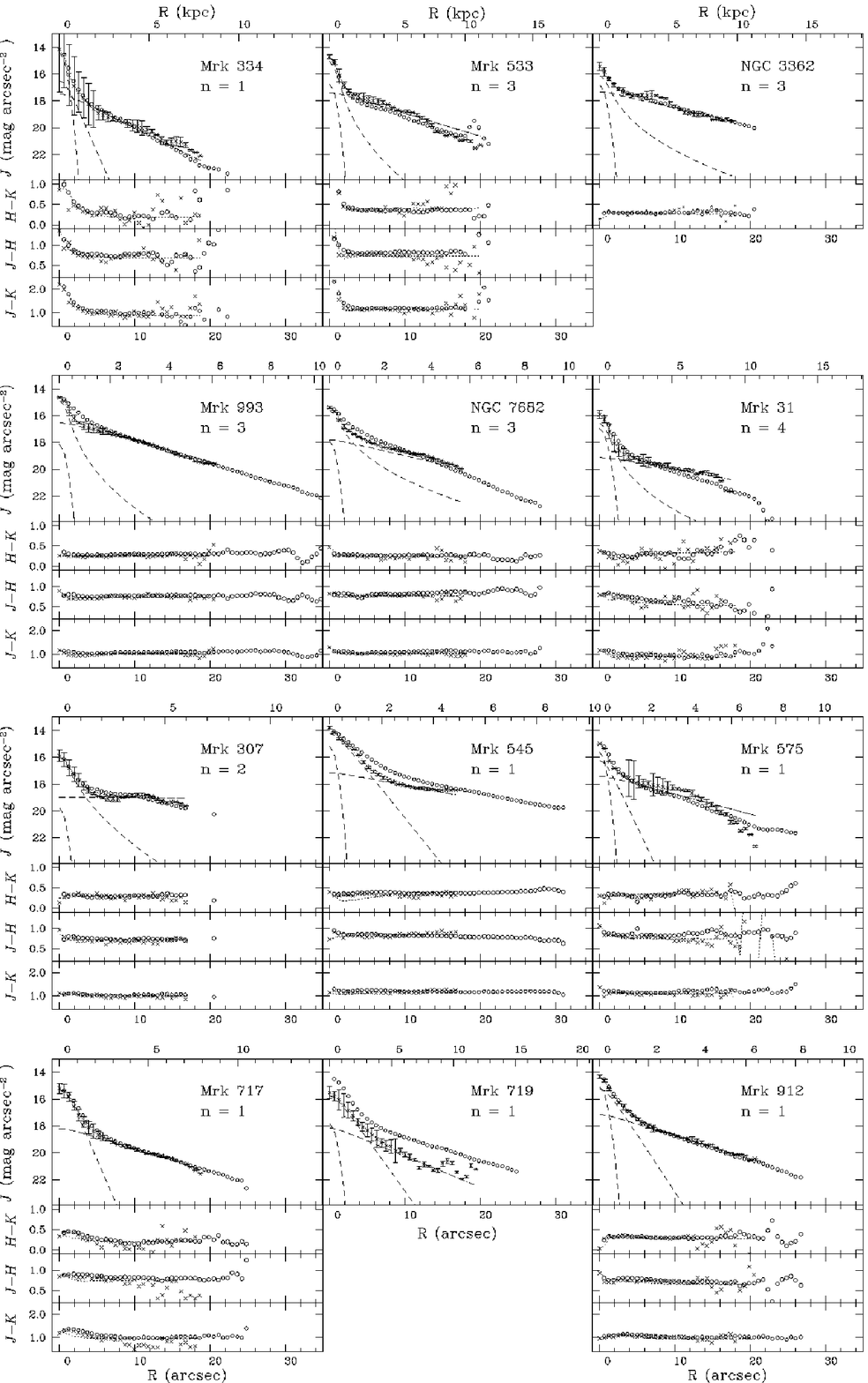,width=13cm}}
\medskip
\centerline{\bf Fig. 1b}
\end{figure*}

\centerline{\epsfig{figure=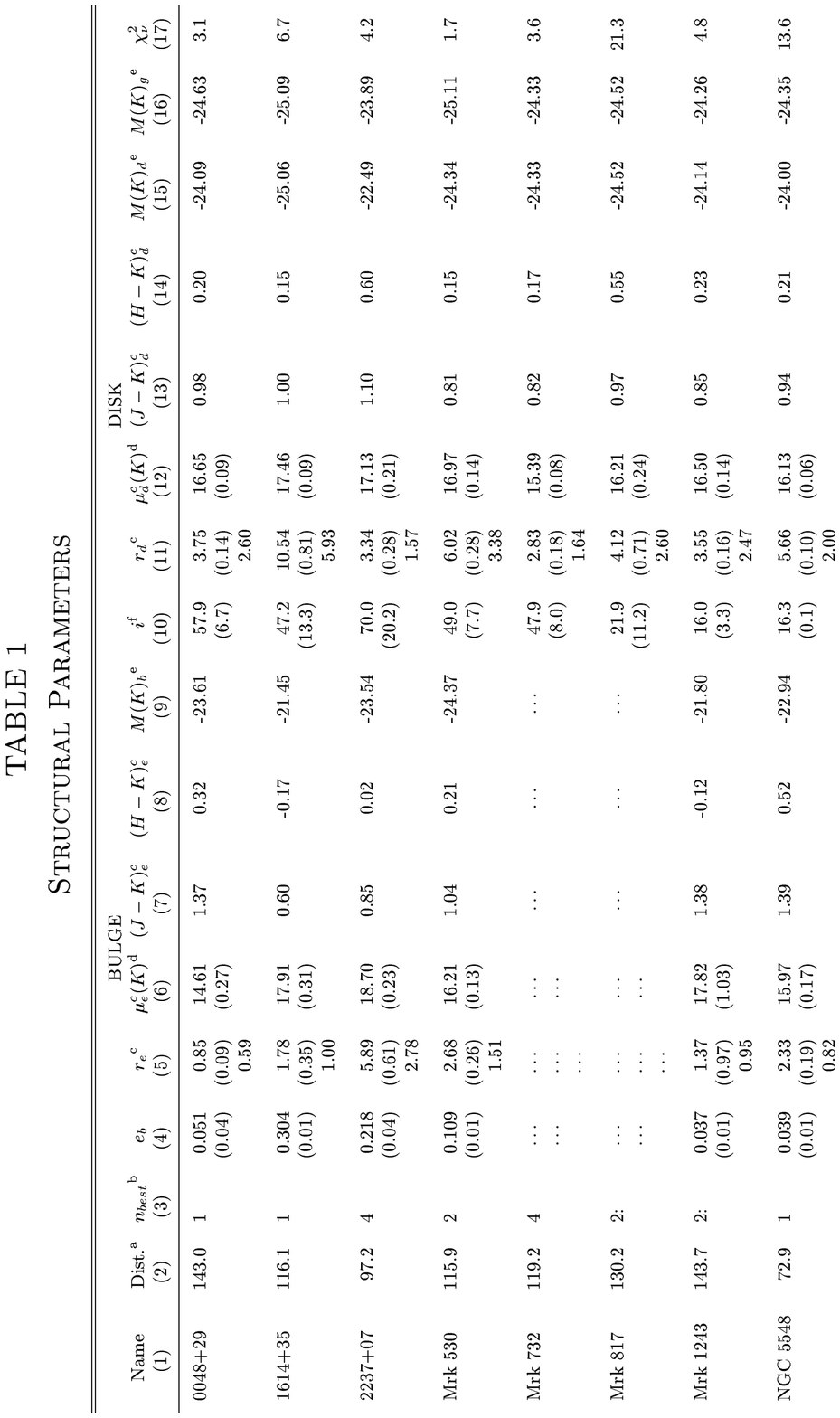,height=25cm,angle=180}}
\centerline{\epsfig{figure=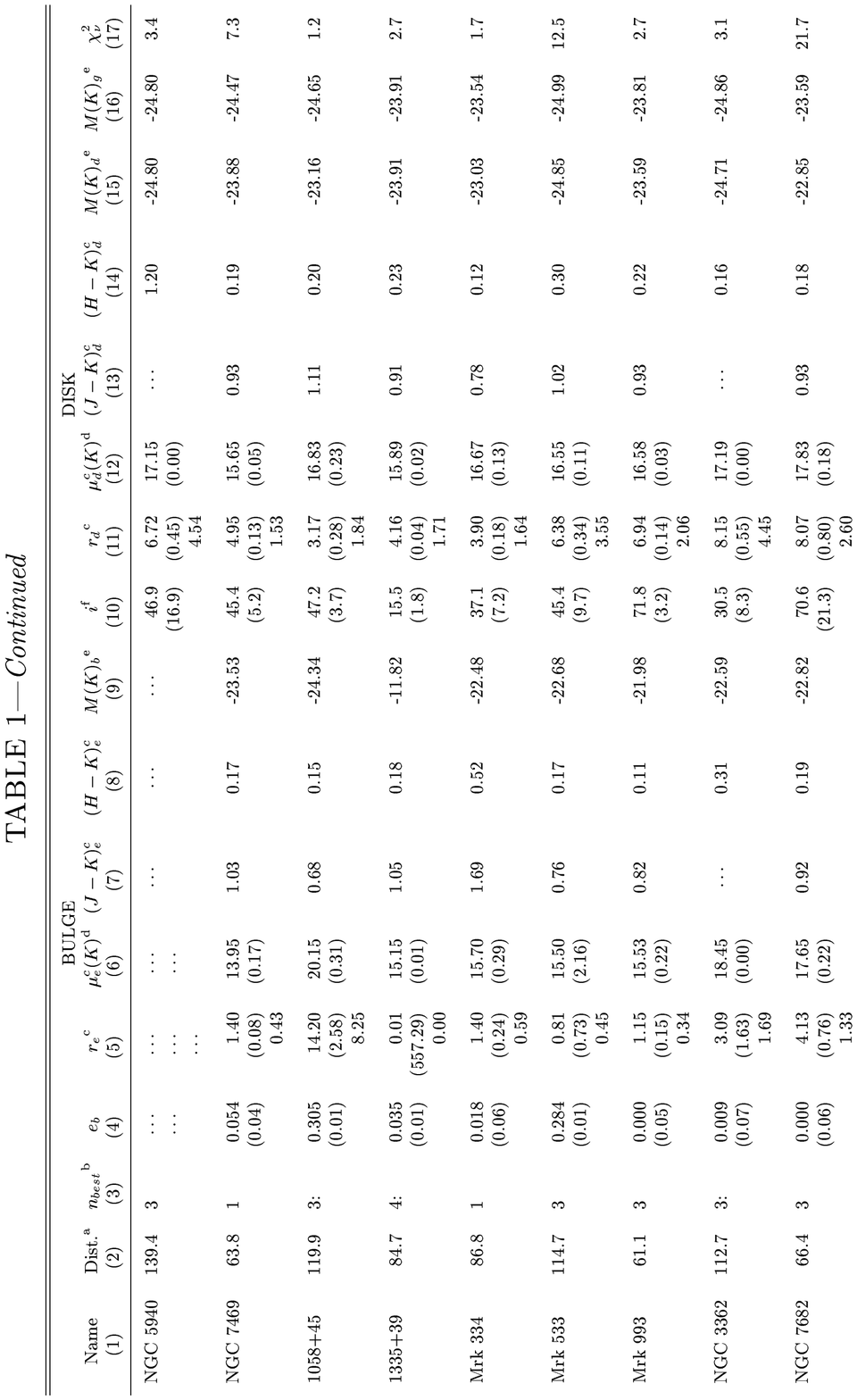,height=25cm,angle=180}}
\centerline{\epsfig{figure=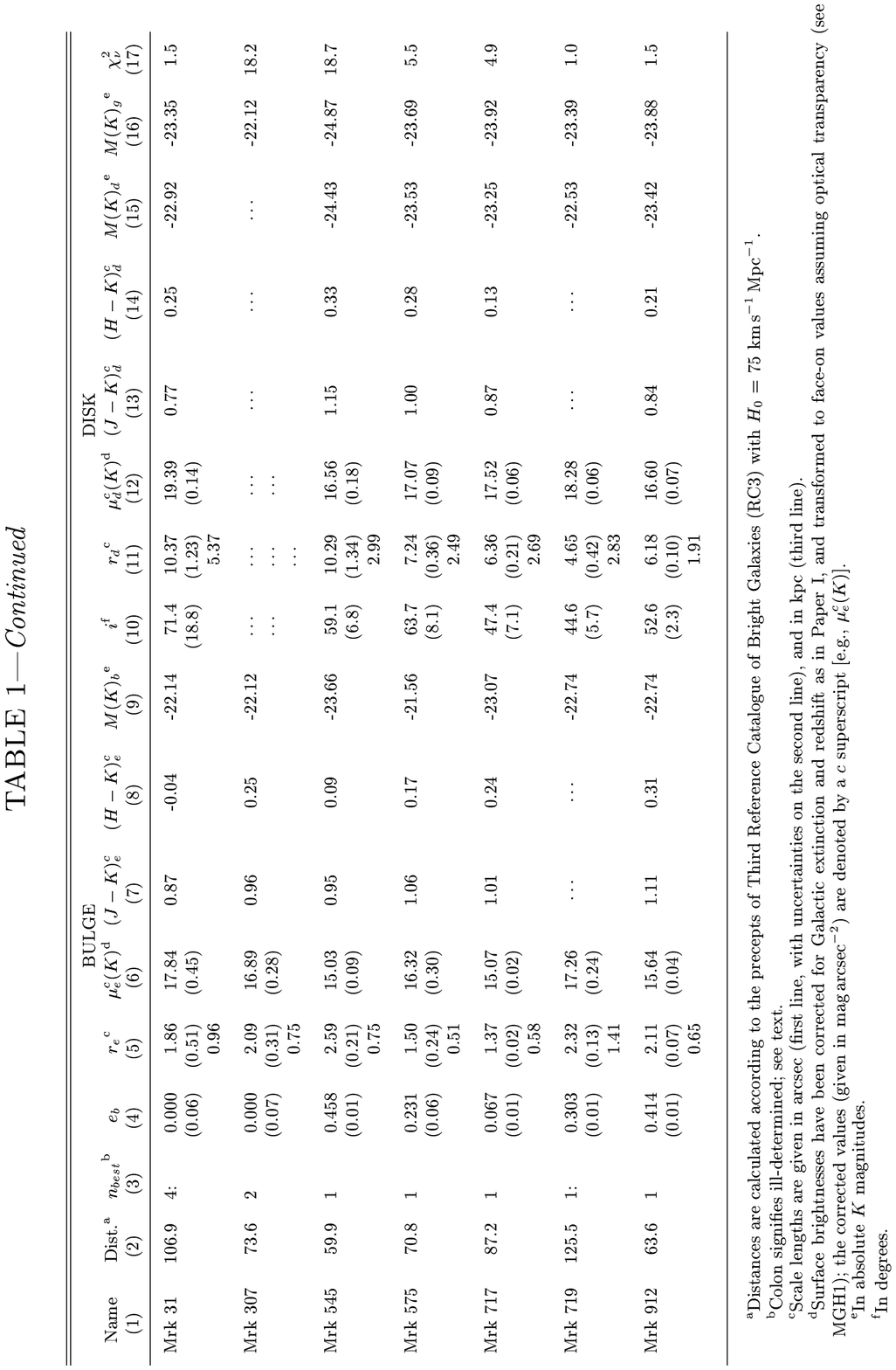,height=25cm,angle=180}}

\end{document}